\documentclass{iopart}
\usepackage{graphicx}
\usepackage{float}

\begin{document}
  \bibliographystyle{unsrt}
  \title [Jammed state characterization of RSA of segments of two lengths]{Jammed state characterization of the random sequential adsorption of segments of two lengths on a line}

  \author{N. A. M. Ara{\'u}jo}
  \address{Computational Physics for Engineering Materials, IfB, ETH Z{\"u}rich, Schafmattstr. 6, 8093 Z{\"u}rich, Switzerland}
  \author{A. Cadilhe}
    \address{GCEP-Centro de F{\'i}sica da Universidade do Minho, 4710-057 Braga, Portugal and T-1 Group, Theoretical Division, MS B-268, Los Alamos National Laboratory, Los Alamos, NM 87545, USA}
    \ead{\mailto{cadilhe@lanl.gov}}

  \begin{abstract}
    We characterize the jammed state structure of the random sequential adsorption of segments of two different sizes on a line.
    To this end, we define the size ratio as a dimensionless quantity measuring the length of the large segments in terms of the smaller ones.
    We introduce a truncated exponential as an {\it ansatz} for the probability distribution function of the interparticle distance at the jammed state and use it to reckon the first four cumulants of the probability distribution function.
    The sole free parameter present in the various analytical expressions is tied to the mean interparticle distance from Monte Carlo simulations, while the remaining three cumulants are computed without any free parameter and compared to Monte Carlo results.
    We find that the proposed {\it ansatz} provides results in good qualitative agreement with Monte Carlo ones.
  \end{abstract}
  \maketitle

  \section*{Introduction}\label{Intro}
    The random sequential adsorption model has been used as a paradigm to study irreversible phenomena like deposition and adsorption of particles onto a substrate in systems where adsorbed particles do not significantly diffuse within the experimental time scale \cite{Bartelt91b,Privman00a,Evans93,Privman94,Privman97,Cadilhe07,Araujo08}.
    Nonetheless, the random sequential adsorption model has also been applied to the modeling of apparently dissimilar systems like the cases of the classical works on intramolecular reactions taking place in polymers \cite{Flory39,Gonzalez74} and the famous car parking problem \cite{Renyi58,Renyi63}.
    The random sequential adsorption model has been also applied to study the irreversible adsorption of colloids \cite{Onoda86,Privman91b} and, more generally, the deposition of extended objects \cite{Ballani06,Gromenko09a,Gromenko09b,Subashiev07a,Subashiev07b}.
    Recently, interest has shifted towards the study of the effect of adsorbing particles of different sizes, with research becoming focused on the study of binary mixtures of particles undergoing either {\it competitive} adsorption or {\it preadsorption}.
    Regarding the latter case, particles adsorb during two subsequent stages with each of the particle species attempting adsorption at a time \cite{Lee96,Cadilhe04,Dorsogna05,Weronski05,Kondrat06}, while during {\it competitive} adsorption both particle species attempt adsorption simultaneously \cite{Subashiev07a,Subashiev07b,Cadilhe04,Bartelt91a,Bonnier01a,Hassan01,Hassan02,Araujo06}.

    In general, several physical properties are related to the interparticle distance, so proper characterization of the corresponding probability distribution function is fundamental.
    The interest stems from the theoretical perspective in gaining insight on the functional dependence of the interparticle distribution even when an analytical solution is not available.
    From the experimental perspective, the availability of the interparticle distribution function provides for controlled analysis.
    In the present work we focus on the {\it competitive} adsorption of segments of two different sizes on a line and characterize various interparticle distribution functions with moments up to the fourth order.
    We propose a novel approach to characterize the gap-size distribution functions\footnote{This is a term that will be interchangeably used to represent the interparticle distribution function.} based on the systematic use of their cumulants.
    Comparing the semi-analytical results using an {\it ansatz} with recently reported Monte Carlo simulation results \cite{Araujo06}, we find good qualitative agreement.

    The paper is organized as follows: In the next section, we start introducing the motivation, some terminology, and theory.
    In Sec.~\ref{Distrib} we introduce the {\it ansatz} and we compute the various cumulants.
    We present our results and final remarks in Secs.~\ref{Res}~and~\ref{concl}.

  \section{Motivation and theory}\label{Model}

    We consider the competitive adsorption of two size segments on the line, namely short segments, which we denote as $A$ segments, and long ones as $B$.
    The model corresponds to an extension of the continuum random sequential adsorption model \cite{Privman00a,Evans93,Cadilhe07,Bonnier01a,Hassan02,Privman00b}, where a segment successfully adsorbs if it does not overlap a previously adsorbed one, therefore mimicking an excluded volume, short-range interaction.
    The model has been studied both analytically \cite{Bonnier01a,Hassan02} and by Monte Carlo simulations \cite{Cadilhe07,Hassan02,Araujo06}.
    Without loss of generality, we rescale the size of the system by the length of short segments, so that the length of the smaller segments becomes unity while that of the longer segments is $R$.
    We term as the size ratio the quotient of the lengths of the larger to the smaller segments, which equals $R$ (the length of the larger segments in dimensionless units).
    In general, one can have both segment types to arrive at different rates per unit length and per unit time, which represents the incident particle flux in the one-dimensional case.
    In the present work, we take the particular case of equal incident fluxes for both types of segments, i.e., on average, equal number of segments attempt deposition per unit time and per unit length.
    Since adsorption of segments is irreversible, i.e., in physical systems where surface mobility and detachment upon adsorption can be safely neglected, it leads to a jammed state, where no segment can fit into any of the available interstitial spaces.
    The present model has been studied in the literature in various limiting cases.
    For example, for $R= 1$ it corresponds to the R{\'e}nyi or car problem result \cite{Bartelt91b,Privman00a,Evans93,Privman94,Privman97,Cadilhe07,Araujo08,Flory39,Gonzalez74,Renyi58,Renyi63,Bonnier01a,Hassan02,Bartelt91c}.
    The case of $R \to \infty$ has been previously discussed \cite{Araujo06}.
    The limit of $R \to 0$ leads to kinetics of pointlike particles or fragmentation kinetics \cite{Bonnier01a,Hassan01,Bartelt91c}.
    Though we limit ourselves to size ratios $R \ge 1$, the pointlike kinetics corresponds to the limit of $R \to \infty$, followed by a rescaling of the linear dimensions by the length of the large particle.

    To properly characterize the jammed state, we start by introducing some definitions and relations.
    The basic quantity to characterize is the probability distribution of the interparticle distance defined as
    \begin{equation}
      \hspace*{-1.0cm}P_{\emptyset}(x)\mbox{d}x= \frac{\ \!\mbox{Number of empty intervals of size }x\mbox{ within }]x, x + \mbox{d}x[\ \!}{\mbox{Total number of empty intervals}}\mbox{d}x.
    \end{equation}
    Thus, $P_{\emptyset}(x)\mbox{d}x$ at the jammed state has the property
    \begin{equation}\label{normalization}
      \int_0^1 P_{\emptyset}(x)\mbox{d}x=1.
    \end{equation}

    Discriminating all gap types in terms of pairs of consecutively adsorbed segments provides four distinct cases, namely $AA$, $AB$, $BA$, and $BB$.
    Density distribution functions $P_b(x)$, with $b\in \{AA, AB, BA, BB\}$, i.e., for each of the gap types, are defined, from which the following relation is obeyed
    \begin{equation}\label{discrimination}
      P_{\emptyset} (x)=P_{AA} (x) + 2P_{AB} (x) + P_{BB} (x),
    \end{equation}
    where the equality $P_{AB} (x)= P_{BA} (x)$ was used.
    Note from equations (\ref{normalization}) and (\ref{discrimination}) that the $P_b (x)$ are not normalized.
    Keeping in mind the above definitions, it is now straightforward to reckon moments of any order of the gap-size distribution functions, defined by
    \begin{equation}\label{moments}
      \langle x^n\rangle _a= \frac{\int_0^1 x^nP_a (x)\mbox{d}x}{\int_0^1P_a (x)\mbox{d}x},
    \end{equation}
    with $a\in \{\emptyset, AA, AB, BA, BB\}$ and $n= 0$, $1$, $2$, \dots, though in the present work, we restrict ourselves to moments with $n \le 4$.
    Using equation (\ref{discrimination}) one can relate the moments of the gap-size distributions with the corresponding moment of the global gap-size distribution function given by equation (\ref{moments}) yielding
    \begin{eqnarray}
      {\langle x^n\rangle}_{\emptyset}&=&{\langle x^n \rangle}_{AA}\int_0^1P_{AA} (x)\mbox{d}x \nonumber \\
      &&\quad+2{\langle x^n\rangle}_{AB}\int_0^1P_{AB} (x)\mbox{d}x \nonumber \\
      &&\quad+{\langle x^n\rangle}_{BB}\int_0^1P_{BB} (x)\mbox{d}x,
    \end{eqnarray}
    again, defined for all values of $n=$ $0$, $1$, $2$, $\dots$.
    We also compute the cumulants, $\kappa_m^a$, of the gap-size distributions defined as,
    \begin{equation}\label{cumulants}
      \ln G_a (k)= \sum_{m= 1}^\infty\frac{(ik)^m}{m!}\kappa_m^a,
    \end{equation}
    where $G_a (k)$ is the so called characteristic function defined by \cite{Kampen92},
    \begin{equation}\label{characteristic}
      G_a (k)\equiv {\langle e^{ikx}\rangle}_a= \frac{\int_0^1e^{ikx}P_a (x)\mbox{d}x}{\int_0^1P_a (x)\mbox{d}x}.
    \end{equation}
    Therefore, from equations (\ref{moments}), (\ref{cumulants}), and (\ref{characteristic}), one derives the first four cumulants as
    \begin{equation}\label{cumulant_k1}
      \kappa_1^a= {\langle x\rangle}_a,
    \end{equation}
    \begin{equation}\label{cumulant_k2}
      \kappa_2^a= \langle x^2\rangle _a-\langle x\rangle _a^2,
    \end{equation}
    \begin{equation}\label{cumulant_k3}
      \kappa_3^a= \langle x^3\rangle _a-3\langle x^2\rangle _a \langle x\rangle _a+2\langle x\rangle _a^3,
    \end{equation}
    \begin{equation}\label{cumulant_k4}
      \kappa_4^a= \langle x^4\rangle _a-4\langle x^3\rangle _a\langle x\rangle _a-3\langle x^2\rangle _a^2+12\langle x^2\rangle _a\langle x\rangle _a^2-6\langle x\rangle _a^4,
    \end{equation}
    where $\kappa_1$ is just the mean value and $\kappa_2$ is the variance.
    While the first and second cumulants have well-established meanings, those of the third and fourth order cumulants are put in terms of the skewness,
    \begin{equation}\label{skewness}
      S_a= \frac{\kappa_3^a}{\left(\kappa_2^a\right)^{3/2}},
    \end{equation}
    and kurtosis,
    \begin{equation}\label{kurtosis}
      K_a= \frac{\kappa_4^a}{\left(\kappa_2^a\right)^2},
    \end{equation}
    respectively \cite{Kampen92}, in order to facilitate their interpretation.
    For later convenience, we also take the opportunity to define the dispersion, $\sigma_a\equiv \sqrt{\kappa_2^a}$, used in the analysis of the results provided in Sec.~\ref{Res}.

  \section{The gap-size distribution functions}\label{Distrib}

    Since the probability that a segment is deposited in a given gap is proportional to the {\it effective} size of the gap, the probability distribution describing the interparticle distance should be reasonably approximated by an exponential distribution.
    The {\it effective} gap size is defined as the difference between the actual size of the gap and the size of the segment attempting adsorption, so it represents the actual length available for deposition.
    In the jamming state, the state where no more particles can be deposited, the probability of having a gap of length larger than one is zero.
    So, we propose as the {\it ansatz} for the probability distribution function of the interparticle distance in the jammed state, an exponential function truncated to the interval $[0, 1[$,
    \begin{equation}\label{Prob}
      P_a (x)= \left\{
                 \begin{array}{ll}
                   \!\!A_a (\alpha_a) e^{-\alpha_a x}, & 0 \le x < 1 \\
                   \!\!0, & x > 1 \\
                   \end{array}
               \right.,
    \end{equation}
    where $\alpha_a$ is the single parameter used to characterize the distribution function of the interparticle distance.
    Since $P_{\emptyset}(x)$ is normalized (eq.~(\ref{normalization})), one has
    \begin{equation}\label{norm.factor}
      A^{-1}_{\emptyset}(\alpha_{\emptyset})= \frac{1 - e^{-\alpha_{\emptyset}}}{\alpha_{\emptyset}}= 1.
    \end{equation}

    Some considerations are due before proceeding to the derivation of the various cumulants.
    The idea of using the proposed {\it ansatz} (eq.~(\ref{Prob})) is to describe the jammed state.
    Though a truncated-exponential {\it ansatz} best describes uncorrelated deposition events, it allows us to assess the presence of correlated deposition events at the jammed state.
    The description of the full kinetics towards the jammed state would not be possible with the proposed {\it ansatz}.
    In particular, during the adsorption process, the upper limit of the integral in eq.~(\ref{normalization}) would have to be extended to infinity.
    Moreover, at early times, the kinetics of deposition is uncorrelated and an exponential distribution represents a better fit \cite{Cadilhe07}.
    Thus, we are interested in the jammed state when systematic elimination, by the adsorption of particles, of all gaps of length $x \ge 1$ have taken place.

    From eq.~(\ref{moments}) the various moments of the gap-size distributions in eq.~(\ref{Prob}) are calculated.
    The first moment equals the first-order cumulant (eq.~(\ref{cumulant_k1})),
    \begin{equation}\label{moment_m1}
      \kappa_1^a\equiv\langle x\rangle_a= \frac{1}{1-e^{\alpha_a}}+\frac{1}{\alpha_a}.
    \end{equation}
    The second moment is
    \begin{equation}\label{moment_m2}
      {\langle x^2\rangle}_a= \alpha_a^{-2}\left[2+\frac{\alpha_a (2+\alpha_a)}{1-e^{\alpha_a}}\right],
    \end{equation}
    and the dispersion is
    \begin{equation}\label{cumulant_teo_k2}
      \sigma_a= \sqrt{\frac{1}{\alpha_a^2}+\frac{1}{2-2 \cosh{\left(\alpha_a\right)}}}.
    \end{equation}

    \begin{figure}[floatfix]
      \includegraphics[width=8.0cm]{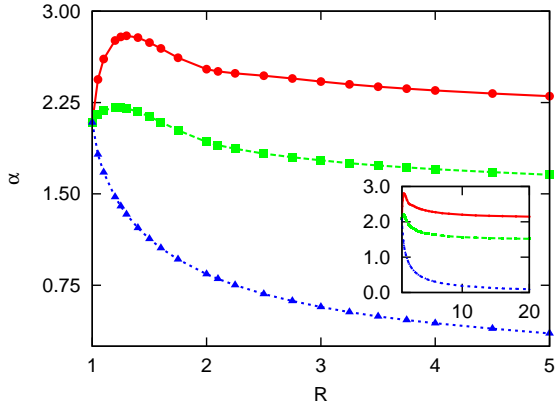}
      \caption
      { \label{fig:1}Plot of $\alpha_a$, with $a\in\left\{AA, AB, BB\right\}$, as a function of size ratio computed from first moment of Monte Carlo simulations \cite{Araujo06}.
        Each of the $AA\mbox{-}$, $AB\mbox{-}$, and $BB$-gap types are represented by circles, squares, and triangles, respectively.
      }
    \end{figure}

    To better characterize the distribution function, as previously referred to, one resorts to the calculation of higher moments.
    The third- and fourth-order cumulants are reckoned in terms of the moments and used for the definition of the skewness and kurtosis, which are quantities with a more intuitive meaning.
    The skewness yields information about the asymmetry of the distribution, while the kurtosis provides information about the peakness or flatness of the distribution function.
    For completeness, we provide the relations for the third and fourth moments,
    \begin{equation}\label{moment_m3}
      {\langle x^3\rangle}_a= {\alpha_a}^{-3}\left[6 + {\frac{\alpha_a(6 + \alpha_a(3 + \alpha_a))}{1 - e^{\alpha_a}}}\right]
    \end{equation}
    and
    \begin{equation}\label{moment_m4}
      {\langle x^4\rangle}_a= {\alpha_a}^{-4}\left[24 + \frac{\alpha_a(24 + \alpha_a(12 + \alpha_a (4 + \alpha_a)))}{1 - e^{\alpha_a}}\right].
    \end{equation}
    After some tedious, though straightforward, algebra we obtain the third and fourth cumulants,
    \begin{equation}\label{cumulant_teo_k3}
      \kappa_3^a= \frac{2}{\alpha_a^3} - \frac{1}{4} \coth{\left(\frac{\alpha_a}{2}\right)}\mbox{csch}^2{\left(\frac{\alpha_a}{2}\right)}
    \end{equation}
    and
    \begin{equation}\label{cumulant_teo_k4}
      \kappa_4^a= \frac{6}{\alpha_a^4} - \frac{1}{8} (2 + \cosh\left(\alpha_a\right)) \mbox{csch}^4{\left(\frac{\alpha_a}{2}\right)}.
    \end{equation}

    \begin{figure}[floatfix]
      \includegraphics[width=11.5cm]{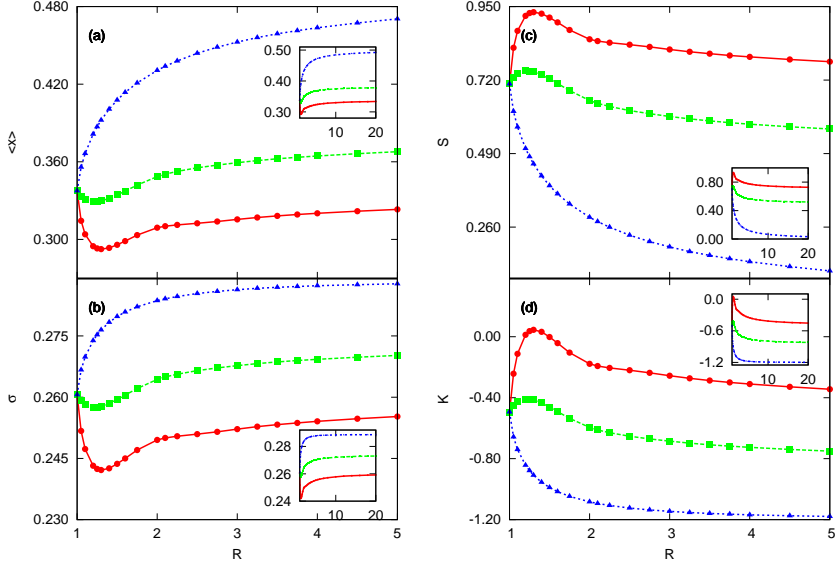}
      \caption{\label{fig:2}Plots involving cumulants up to the fourth order of the gap-size distribution functions for each gap type as a function of the size ratio, according to equations (\ref{moment_m1}), (\ref{cumulant_teo_k2}), (\ref{skewness_teo}), and~(\ref{kurtosis_teo}).
        a) Average  and
        b) dispersion of the distance between pairs of segments.
        c) Skewness.
        d) Kurtosis.
      The legend convention follows that presented in Fig.~\ref{fig:1}.
      }
    \end{figure}

    From eqs.~(\ref{skewness}) and~(\ref{kurtosis}), and using eqs.~(\ref{cumulant_teo_k2}), (\ref{cumulant_teo_k3}), and~(\ref{cumulant_teo_k4}), we obtain the skewness
    \begin{equation}\label{skewness_teo}
      S_a=\frac{4-4 \cosh\left(\alpha_a\right)+\alpha_a^3 \coth\left(\frac{\alpha_a}{2}\right)}{\alpha_a \sqrt{\frac{1}{\alpha_a^2}+\frac{1}{2-2 \cosh\left(\alpha_a\right)}} (2+\alpha_a^2-2 \cosh\left(\alpha_a\right))},
    \end{equation}
    and the kurtosis
    \begin{equation}\label{kurtosis_teo}
      K_a=-2\frac{(24+\alpha_a^4) \cosh\left(\alpha_a\right)+2 (-9+\alpha_a^4-3 \cosh\left(2 \alpha_a\right))}{(2+\alpha_a^2-2 \cosh\left(\alpha_a\right))^2},
    \end{equation}
    for the truncated exponential distribution function defined in eq.~(\ref{Prob}).

  \section{Results and Discussion}\label{Res}

    In a previous work \cite{Araujo06}, we have performed an extensive Monte Carlo simulation study of the present random sequential model.
    Here, we utilize results from that work, specifically from the first moment as defined in eq.~(\ref{moment_m1}), to determine the values of $\alpha$ for all gap types and size ratios ranging from $1$ to $20$.
    Figure~\ref{fig:1} shows the computed values of $\alpha$ as a function of the size ratio.
    Heretoforth, the values of the size ratio match those of the Monte Carlo simulations to simplify comparison~\cite{Araujo06}.
    For the $BB$ gap type a monotonic decrease of $\alpha$ with the size ratio is observed.
    The $AA$ and $AB$ gap types are characterized by maxima for values of the size ratio at $1.30\pm0.05$ and $1.20\pm0.05$, respectively.

    To characterize the system, in Fig.~\ref{fig:2} the average distance, dispersion, skewness, and kurtosis are computed, respectively, with eqs.~(\ref{moment_m1}), (\ref{cumulant_teo_k2}), (\ref{skewness_teo}), and~(\ref{kurtosis_teo}) as a function of the size ratio.
    For the $AA$ and $AB$ gap types, a non-monotonic behavior for all these quantities is observed.
    For the BB gap, the average distance and dispersion increase with the size ratio as shown in Fig.~\ref{fig:2}(a) and~(b), while the skewness and kurtosis decrease monotonically as shown in parts~(c) and~(d) of the same figure.
    For the average distance and dispersion (Fig.~\ref{fig:2}(a)~and~(b)) there is a minimum for the $AA$ gap at a value of the size ratio of $1.30\pm0.05$ and of $1.20\pm0.05$ for the $AB$ gap.
    For the skewness and kurtosis (Fig.~\ref{fig:2}(c)~and~(d)) there is a maximum for the $AA$ gap at $1.30\pm0.05$ and at $1.20\pm0.05$ for $AB$ gap.
    The values of the maxima and minima for the $AA$ and $AB$ gap types coincide with the minimum values of $\alpha$ as a function of the size ratio.

    \begin{figure}[floatfix]
      \includegraphics[width=12.0cm]{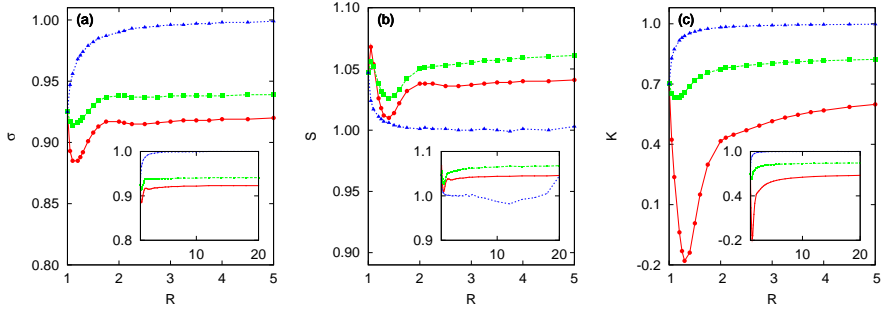}
      \caption{\label{fig:3}Plots involving the ratio of cumulants of the gap-size distribution functions of the same order.
        The plots provide a direct comparison of the results from the truncated exponential {\it ansatz} to those from Monte Carlo simulations as explained in the text.
          a) Dispersion.
          b) Skewness.
          c) Kurtosis.
        The legend convention follows that of Fig.~\ref{fig:1}.
      }
    \end{figure}

    We compare results using the truncated exponential {\it ansatz} with simulational ones found in Ref.~\cite{Araujo06}.
    All computed quantities possess the same qualitative behavior as that obtained from an extensive Monte Carlo study.
    The exception regards the values of the dispersion at sizes ratios near unit.
    In this situation, streaks of alternating $A$ and $B$ segments are common, since two consecutive $BB$ segments might still have a gap of length larger than unit, where a single $A$ segment can fit in.
    In fact, at size ratios lower than $1.55$, the population of $BB$ gaps remains significant, as compared with the $AA$ gap one.
    However such streaks leading to gaps of size near unit, which we denoted as {\it snug fits} in Ref.~\cite{Araujo06}, are not possible to be described by a simple truncated exponential as the latter does not account for influences due to the history of several segment deposition events.
    The ratios of the values of the cumulants of the same order from the truncated exponential {\it ansatz} to those from Monte Carlo simulations are show in Fig.~\ref{fig:3}.
    The $BB$ gap shows better compliance with proposed {\it ansatz}.
    In a sense, this result is not entirely surprising, since the larger segments are deposited at early times.
    Consequently, there will be less interaction with previously deposited segments.
    At late times, i.e., in the asymptotic regime, the $A$ segments will fill the gaps left out by the $B$ segments, particularly as the size ratio increases.
    Improved compliance to the truncated exponential is also observed for the $AA$ and $AB$ gap types with increasing values of the size ratio for similar reasons to those of the $BB$ gaps.
    As the size ratio increases, the $B$ segments have a sharper decay in the rate of adsorption due to single $A$ segments blocking their adsorption.
    Consecutive $B$ segments also leave larger spacing between them, on average.
    Consequently, the adsorption of $A$ segments follows more closely the argument in favor of a truncated exponential presented above.
    At size ratios close to one, the results provided by the {\it ansatz} deviate more, since the populations of the various gap types are nearly equilibrated and one expects the equations describing these gap types to be more {\it coupled}, i.e., to be more dependent on multisegment deposition events.

  \section{Final remarks}\label{concl}
    In short, we introduced a novel approach with potential for a semi-quantitative analysis of models where the interparticle distribution function of the distances is of relevance.
    Specifically, the method simply uses a single moment to obtain information for the other three.
    However, it requires some prior knowledge of what can constitute a good {\it ansatz}.
    We present a coherent study of the dependence on the size ratio up to the fourth moment.
    We propose a truncated exponential function for the gap-size distribution and compute the mean gap size, dispersion, skewness, and kurtosis.
    Qualitatively, the results for the proposed {\it ansatz} are in good agreement with the results obtained from simulational work.
    However, the dispersion of AB gap type at small values of size ratio, the exponential function does not reproduce the behavior of the distribution function, which {\it can be explained by {\it snug fit} events}.
    Finally, we point out that the present approach can be applied to a broader range of models involving the irreversible deposition of particles on a line.

  \begin{ack}
    This research has been funded by a Funda{\c c}{\~a}o para a Ci{\^e}ncia e a Tecnologia research grant and Search (Services and Advanced Research Computing with HTC/HPC clusters) (under contract CONC-REEQ/443/2001).
    One of us, N.~A., thanks Funda{\c c}{\~a}o para a Ci{\^e}ncia e a Tecnologia for a Ph.D.\ fellowship (SFRH/BD/17467/2004).
    A.~C. wants also to thank both Funda{\c c}{\~a}o para a Ci{\^e}ncia e a Tecnologia (SFRH/BPD/34375/2007) and Funda{\c c}{\~a}o Calouste Gulbenkian for fellowships to visit Los Alamos National Laboratory.
    A.~C. also acknowledges the warm hospitality of the T-1~Group at Los Alamos National Laboratory.
    Finally, we want to thank suggestions and comments on earlier versions of the manuscript by N.~Henson, V.~Privman, and C.~Reichhardt.
  \end{ack}

  \section*{References}
  \bibliography{rsa1d,rsa1dextra}
\end{document}